\documentclass[12pt]{article}
\usepackage{graphicx}
\usepackage{epsfig}
\usepackage{mathtext}
\newcommand{\beq}{\begin{equation}}
\newcommand{\eeq}{\end{equation}}
\newcommand{\beqn}{\begin{eqnarray}}
\newcommand{\eeqn}{\end{eqnarray}}
\begin{document} 
 
\title{\textbf{WHAT ARE THE NEUTRINO MASSES. MIXING}} 
\author{V.P. Efrosinin\\
Institute for Nuclear Research\\
Moscow 117312, Russia}

\date{}
\renewcommand {\baselinestretch} {1.3}

\maketitle
\begin{abstract}
The possible mechanisms of generation of the neutrino of small masses and
sterile neutrino are considered. In  the supposition of stability of tracks of
a left-right symmetry concerning masses of constituent quarks, the angles
of mixing of quarks are evaluated.
With usage of experimental data on neutrino oscillation parameters the neutrino
masses on order of magnitude are computed.
    
\end{abstract}

One of most difficult for comprehension of properties of weak interactions is
the phenomenon of small neutrino masses.
From experiments on $\textup{H}^3$ $\beta$-decay follows
\cite{bonn,loba}
\begin{eqnarray}
\label{eq:MM1}
m_{\beta} < 2.2~\textup{eV}. 
\end{eqnarray}
From cosmology data the sum of neutrino masses has high bound
\begin{eqnarray}
\label{eq:MM2}
\sum_i m_i \leq (0.4 \div 1.7)~\textup{eV}.
\end{eqnarray}
Besides from experiments on looking up of neutrino oscillation
\cite{aliu,fuku}, the neutrino masses of different heerarchies are clous,
\begin{eqnarray}
\label{eq:MM3}
\Delta m^2_{ji} = m^2_j-m^2_i 
\end{eqnarray}
And as against quarks, the large angles of mixing take place
\begin{eqnarray}
\label{eq:MM4}
sin^2 2\theta_{ji} \sim 1.0. 
\end{eqnarray}
According to the cosmological data, more than one sterile neutrino
mixed with active neutrinos is allowed.
The available experimental data testify, to diverse than for quarks, to the
nature of lepton masses.

As shown in \cite{efro}, at neutrino production with small mass in some
reaction is possible also neutrino production with large mass with
cross section suppressed on $\sim \alpha$ as contrasted to by cross section
of reaction with departure a neutrino of small mass.
It is one of possible mechanisms of appearance sterile neutrinos.
If those will not take place, that it is possible to speak about strong
CP-violation $\sim \alpha$ in weak leptonic top of a charged current.

We are interested by capability of definition of masses of three neutrino
heerarchies with the help of parameters known from experiments of their mixing.
For this purpose we shall consider, in the beginning, relation of angles of
mixing in a weak interaction in six-quark model with the help of ratio of quark
masses. That it is required to us in further presentation. Such problem was
resolved in 
\cite{frit} with usage of the gauge group
$SU^L_2 \times SU^R_2 \times U_1$. We follow to the approach of \cite{frit}
with some changes:

1. We use the standard parametrization \cite{chau}
of the Cabibbo-Kobajashi-Maskawa-matrix $V_{CKM}$
\begin{equation}
V_{CKM}= \left( \begin{array}
{ccl} c_{12}c_{13} & s_{12}c_{13} & s_{13}e^{-i\delta} \\
-s_{12}c_{23}-c_{12}s_{23}s_{13}e^{i\delta} &
c_{12}c_{23}-s_{12}s_{23}s_{13}e^{i\delta}
& s_{23}c_{13} \\
s_{12}s_{23}-c_{12}c_{23}s_{13}e^{i\delta}
& -c_{12}s_{23}-s_{12}c_{23}s_{13}e^{i\delta} & c_{23}c_{13} 
\end{array} \right).
\label{eq:higg}
\end{equation}

2. We are failed from nearings on the first degree of relations $m_s/2m_b$ and
$m_c/2m_t$, and are solved a problem of a diagonalization of a mass matrix
numerically and precisely.

3. We use in calculation of masses of constituent quarks as well as in
\cite{scad,gapo}, insted of masses of current quarks. Confiming such
replacement refs. \cite{efr,efros} testify.

In view of a symmetry of model, of preservation of parity and supposition about
cascade nature of generation of masses, the applicable mass term in a system of
condition of a weak interaction
$(u_0,c_0,t_0)$ was obtained as \cite{frit}:
\begin{equation}
(\overline{u_0c_0t_0})_L
\left( \begin{array}{ccc} 0 & a & 0 \\
a^* & 0 & b \\ 0 & b^* & c \end{array} \right)
\left( \begin{array}{c} u_0 \\ c_0 \\t_0 \end{array} \right)_R
+ L \to R.  
\label{eq:higg1}
\end{equation}
Here $a$, $b$ are complex, $c$ is real number. The similar expression for
a mass term is accepted for $(d_0,s_0,b_0)$ - system (with other parameters
$(a^{\prime}, b^{\prime}, c^{\prime})$). This there was a fair formula for
Cabibbo angle $\theta_{12}$
\begin{eqnarray}
\label{eq:MM7}
\theta_{12}=\arctan(m_d/m_s)^{1/2}-\arctan(m_u/m_c)^{1/2}, 
\end{eqnarray}
obtained in \cite{fri1} for mixing only of two quark hierarchies.

In our calculation of mixing of quarks we have used values of parameters
counted in nonrelativistic model of constituent quarks
\cite{efro,efr1},
\begin{eqnarray}
\label{eq:MM8}
m_u=0.305~\textup{GeV},~m_d=0.305~\textup{GeV},~m_s=0.487~\textup{GeV},
\end{eqnarray}
and also in potential models of high and light mesons \cite{efr2,efr3}
\begin{eqnarray}
\label{eq:MM9}
m_c=1.387~\textup{GeV},~m_b=4.8~\textup{GeV}.
\end{eqnarray}
For a constituent $t$-quark, we have accepted value \cite{yao}
\begin{eqnarray}
\label{eq:MM10}
m_t=173~\textup{GeV}.
\end{eqnarray}

Using values for quark masses
(\ref{eq:MM8})-\ref{eq:MM10}), we receive angles of quark mixing
\begin{eqnarray}
\label{eq:MM11}
\theta_{12}=13^{\circ},~\theta_{23}=-6^{\circ}.
\end{eqnarray}
Let us remark that under the formula (\ref{eq:MM7}) we also receive
\begin{eqnarray}
\label{eq:MM12}
\theta_{12}=13^{\circ},
\end{eqnarray}
that is at mixing of three and two quarks Cabibbo angle a little varies.
From experiment \cite{yao,bona,char} follows
\begin{eqnarray}
\label{eq:MM13}
\theta_{12}=13.14^{\circ} \pm 0.06^{\circ},~
|\theta_{23}|={2.43^{\circ}}^{+0.01^{\circ}}_{-0.05^{\circ}}.
\end{eqnarray}
The value of an angle $\theta_{23}$ in model of the left-right symmetry will
not be agreeed experiment
(\ref{eq:MM13}). In refs. \cite{scad,gapo} the value of this angle is computed
beyond the framework of model, used by us, and better will be agreed experiment.
But we are interested by usage of the given model for propagation on neutrino
mixing, as in experiments the measurements of pair mixing dominate.

We once again reverse attention, that in \cite{frit} in view of
phenomenological limitations $m_t \gg m_c$, $m_b \gg m_s$, there was a fair
formula (\ref{eq:MM7}) for Cabibbo angle and at mixing of three quark
heerarchies. The masses of light quarks  $u,~d$ are generated by mixing in 
a weak interaction.

Therefore in the supposition that the given model will eligible both in lepton
sector, and according to the formula (\ref{eq:MM7}), it is possible to record
expression for a pair angle of oscillation $\nu_e-\nu_{\mu}$
$\theta_{12}$
\begin{eqnarray}
\label{eq:MM14}
\theta_{12}&=&\arctan(m_1/m_2)^{1/2}-\arctan(m_e/m_{\mu})^{1/2}. 
\end{eqnarray}
Or it is possible written as:
\begin{eqnarray}
\label{eq:MM15}
\theta_{12}&=&\arctan(m_1/m_2)^{1/2}-3.98^{\circ}. 
\end{eqnarray}

The already measured neutrino oscillation parameters from global analysis of
$\nu$ data, including solar, atmospheric, reactor and accelerator experiments
\cite{schw},
\begin{eqnarray}
\label{eq:MM16}
\Delta m^2_{12}~&=&~7.6~~(7.3-8.1)~~[10^{-5}~\textup{eV}^2], \nonumber\\
\Delta m^2_{23}~&=&~2.4~~(2.1-2.7)~~[10^{-3}~\textup{eV}^2], \nonumber\\
\sin^2\theta_{12}~&=&~0.32~~(0.28-0.37), \nonumber\\
\sin^2\theta_{23}~&=&~0.50~~(0.38-0.63), \nonumber\\
\sin^2\theta_{13}~&=&~0.007~~(\leq~0.033) 
\end{eqnarray}
will be determined with higher accuracy.

We receive average value
\begin{eqnarray}
\label{eq:MM17}
\sin^2\theta_{12}~=~0.32,~~\theta_{12}~=~34.45^{\circ}
\end{eqnarray}
from (\ref{eq:MM16}). Allowing uncertainty of used model we compute neutrino
masses on an order magnitude. Also we receive
\begin{eqnarray}
\label{eq:MM18}
34.45^{\circ}&=&\arctan(m_1/m_2)^{1/2}-3.98^{\circ}. 
\end{eqnarray}
From here we have
\begin{eqnarray}
\label{eq:MM19}
m_1/m_2~=0.63. 
\end{eqnarray}
With usage of values for $\Delta m^2_{12}$ and $\Delta m^2_{23}$
from (\ref{eq:MM16}) is received
\begin{eqnarray}
\label{eq:MM20}
m_1&=&0.0069 \div 0.0073~\textup{eV}, \nonumber\\
m_2&=&0.0110 \div 0.0116~\textup{eV}, \nonumber\\
m_3&=&0.0471 \div 0.0532~\textup{eV},  
\end{eqnarray}
and
\begin{eqnarray}
\label{eq:MM21}
\sum_i m_i \simeq (0.065 \div 0.072)~\textup{eV}.
\end{eqnarray}
The computed value for $\sum_i m_i$ will be agreeed high bound for the sum
of neutrino masses from the cosmological data. But it is necessary to tell,
that it is an estimation of neutrino masses on an order of magnitude and
generally speaking, modelly dependent.
Any reference points this occasion nevertheless are not known yet.

In summary it would be desirable to pay attention to that circumstance
that the most relevant advances in comprehension of weak coupling are connected
to the non-stationary theory.
Just a solution the problem of $K^0 - \bar{K}^0$ - mixing and neutrino
oscillation. It is difficult to present as though these problems were decided
in the stationary theory of
$S$ - matrix. And here such problem \cite{efro} as
$CP$ - violation in leptonic top of the electroweak theory:
\begin{eqnarray}
\label{eq:M22}
L^{CC}=-\frac{g}{2\sqrt{2}}j^{CC}_{\alpha}W^{\alpha}+э.с., 
\end{eqnarray}
where $g$ is SU(2) gauge coupling constant and current $j^{CC}_{\alpha}$
reads:
\begin{eqnarray}
\label{eq:M23}
j^{CC}_{\alpha}=2\sum_{l=e,\mu,\tau}\bar{\nu}_e\gamma_{\alpha}l_L. 
\end{eqnarray}
The problem directly is connected to the neutrino nature and rather small masses
of these particles.
From the point of view of the stationary theory a neutrino do not take off
in an exited state, as against a charged lepton, and the large 
$CP$ -violation is not watched.
At the same time in top has a place large $CP$ - violation.
But it is not visible in the stationary theory.
At all events looking up a sterile neutrino is rather justified.
Let us remark also that from analysis conducted in \cite{efro} follows
$\sin{\theta}_{13}~\leq~0.007$. So results (\ref{eq:MM16}) for
$\sin^2\theta_{13}$ in the subsequent experiments essentially will change.

\end{document}